\documentclass{mn2e}
\usepackage{graphicx}
\usepackage{amsbsy}
\usepackage{amsmath}
\usepackage{epsf}

\newcommand\apj{ApJ}
\newcommand\mnras{MNRAS}
\newcommand\apjl{ApJ}
\newcommand\apjs{ApJS}

\begin{document}

\title[Too Much Noise about Noise]{Self-Consistent Analysis of OH-Zeeman Observations: Too Much Noise about Noise}

\author[Mouschovias \& Tassis]
{Telemachos Ch. Mouschovias$^{1}$ and Konstantinos Tassis$^{2}$\\
$^1$
Departments of Physics and Astronomy, University of Illinois at Urbana-Champaign, 
1002 W. Green Street, Urbana, IL 61801, USA\\
$^2$
Jet Propulsion Laboratory, California Institute of Technology,
Pasadena, CA 91109, USA}

\maketitle

\label{firstpage}
\begin{abstract}

We had recently re-analyzed in a self-consistent way OH-Zeeman observations in four molecular-cloud envelopes and we had shown that, contrary to claims by Crutcher et al., there is no evidence that the mass-to-flux ratio decreases from the envelopes to the cores of these clouds. The key difference between our data analysis and the earlier one by Crutcher et al. is the relaxation of the overly restrictive assumption made by Crutcher et al, that the magnetic field strength is independent of position in each of the four envelopes. In a more recent paper, Crutcher et al. (1) claim that our analysis is not self-consistent, in that it misses a cosine factor, and (2) present new arguments to support their contention that the magnetic-field strength is indeed independent of position in each of the four envelopes. We show that the claim of the missing cosine factor is false, that the new arguments contain even more serious problems than the Crutcher et al. original data analysis, and we present new observational evidence, independent of the OH-Zeeman data, that suggests significant variations in the magnetic-field strength in the four cloud envelopes.
\end{abstract}


\section{Introduction}

The ambipolar-diffusion theory of molecular-cloud fragmentation and protostar formation predicts mass-to-flux ratios of fragments (or cores) greater than those of their parent clouds (see Fiedler \& Mouschovias 1993, Fig. 9b). Based on this prediction, Cruther, Hakobian \& Troland (2009; hereinafter CHT09) carried out OH Zeeman observations in four molecular cloud envelopes in which the core magnetic field strength was already known. For each cloud, they attempted to measure the envelope magnetic field, $B_{\rm env}$, at four different locations surrounding the core. Of the sixteen measurements only one yielded a 3$\sigma$ detection. The other fifteen yielded only upper limits. In analyzing their data, the authors assumed that (1) the envelope and core magnetic fields of each cloud were in the same direction; and (2) the magnitude of the envelope magnetic field was the same at all four observed locations in each cloud; i.e., that the magnetic field in each cloud envelope was uniform. Hence, the only uncertainties allowed in their analysis were the measurement errors. They concluded that, in all four clouds, the mass-to-flux ratios of the cores were smaller than those of the corresponding envelopes by a factor 0.02 - 0.42, and, therefore, the observations contradict what the authors refer to as ``the idealized ambipolar diffusion theory'' (i.e., the ambipolar diffusion calculations that assume straight-parallel field lines as initial conditions in the parent molecular clouds).

The CHT09 data were re-analyzed by Mouschovias \& Tassis (2009; hereinafter MT09) by relaxing one and only one of the CHT09 assumptions, namely, the one referring to a constant value of the magnetic-field strength in each of the four cloud envelopes. They used a general, likelihood analysis that allows for the possibility that the magnetic-field strength in each cloud envelope has spatial variations -- a possibility suggested by the data themselves and other observations of these four clouds (see \S\,\ref{maps} below). Since most of the measurements were nondetections, MT09 obtained proper upper limits for the relative magnitude ($R$) of the mass-to-flux ratio of each core and that of its envelope. These upper limits were in the range 1.1 - 5.0 -- in sharp contrast with the CHT09 conclusion.

MT09 also pointed out that there are physical reasons for which not only the magnitude, but also the direction of the magnetic field can vary both from a core to its envelope and within an envelope itself. The underlying cause of the expected deformation of the field lines is the motion of cores within a cloud (carrying their field lines with them) and, also, of the cloud relative to the intercloud medium. However, MT09 did {\it not} include in their data analysis any variation of the magnetic field {\it direction} from each core to its envelope for two reasons: (1) There is no reliable way to deduce from the CHT09 data the direction of the magnetic field either in the cores or in the envelopes, a fact that would introduce significant uncertainties in the analysis. (2) MT09 wanted to demonstrate that, when CHT09's overly restrictive assumption of spatially constant value of the field in each envelope is relaxed, as suggested by the data themselves, and proper upper limits are reported for the nondetections, the strong conclusion of CHT09 on the variation of the mass-to-flux ratio from cores to envelopes is shown to be overinflated and completely unjustified.

More recently, Crutcher, Hakobian \& Troland (2010; hereinafter CHT10) responded to the MT09 paper by (1) claiming that the MT09 analysis is not self-consistent, in that a cosine factor is missing, which would account for the different directions of the field in a core and its envelope; and by (2) presenting additional arguments for which the original CHT09 data analysis is proper. As already explained above, the claim about a missing cosine factor is an inaccurate representation of the MT09 analysis, which did not assume a different direction of the core and envelope magnetic field {\it vectors}. We quote from MT09, end of \S\,2:

``In our analysis, we relaxed only {\it one} of the CHT assumptions (that of lack of 
spatial variation of $B_{\rm env}$, which is not consistent with the data). We have 
retained the implicit assumption of similar orientations of the {\it net} 
${\mathbf B}_{\rm env}$ and ${\mathbf B}_{\rm core}$ ({\it vectors}), because the 
data do not suggest any particular relative orientation of the two vectors. 
A more general analysis that would also relax this assumption would increase the 
uncertainties on $R$ (although not on $B_{\rm env}$) and would further part from 
the CHT conclusions.'' 

The main point of the MT09 paper is that the CHT09 data analysis is seriously flawed even if one ignores the theoretically expected possibility that the core and envelope magnetic fields will have different directions. In this paper, we first present observational evidence independent of the CHT09 data, which shows considerable density structure in the four observed envelopes -- thus suggesting field-strength variations as well -- and then show that the new arguments of CHT10 are even more flawed than the original CHT09 analysis; they violate basic rules of logic and scientific reasoning.

\section{Independent Observational Evidence Suggesting $B$-field Variations in the Observed Cloud Envelopes} \label{maps}

\begin{figure*}
\begin{center}
\includegraphics[width=6.5cm]{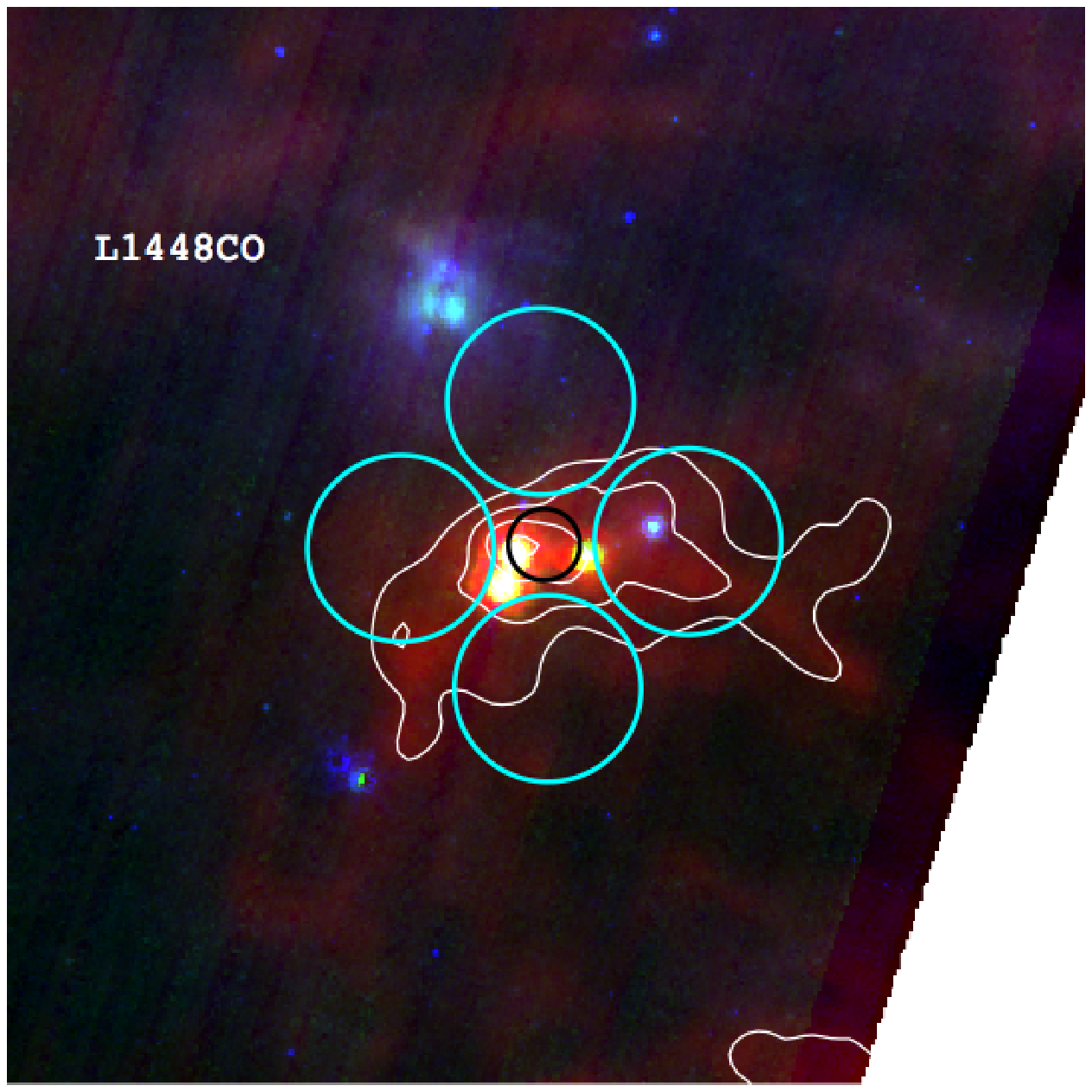}
\includegraphics[width=6.5cm]{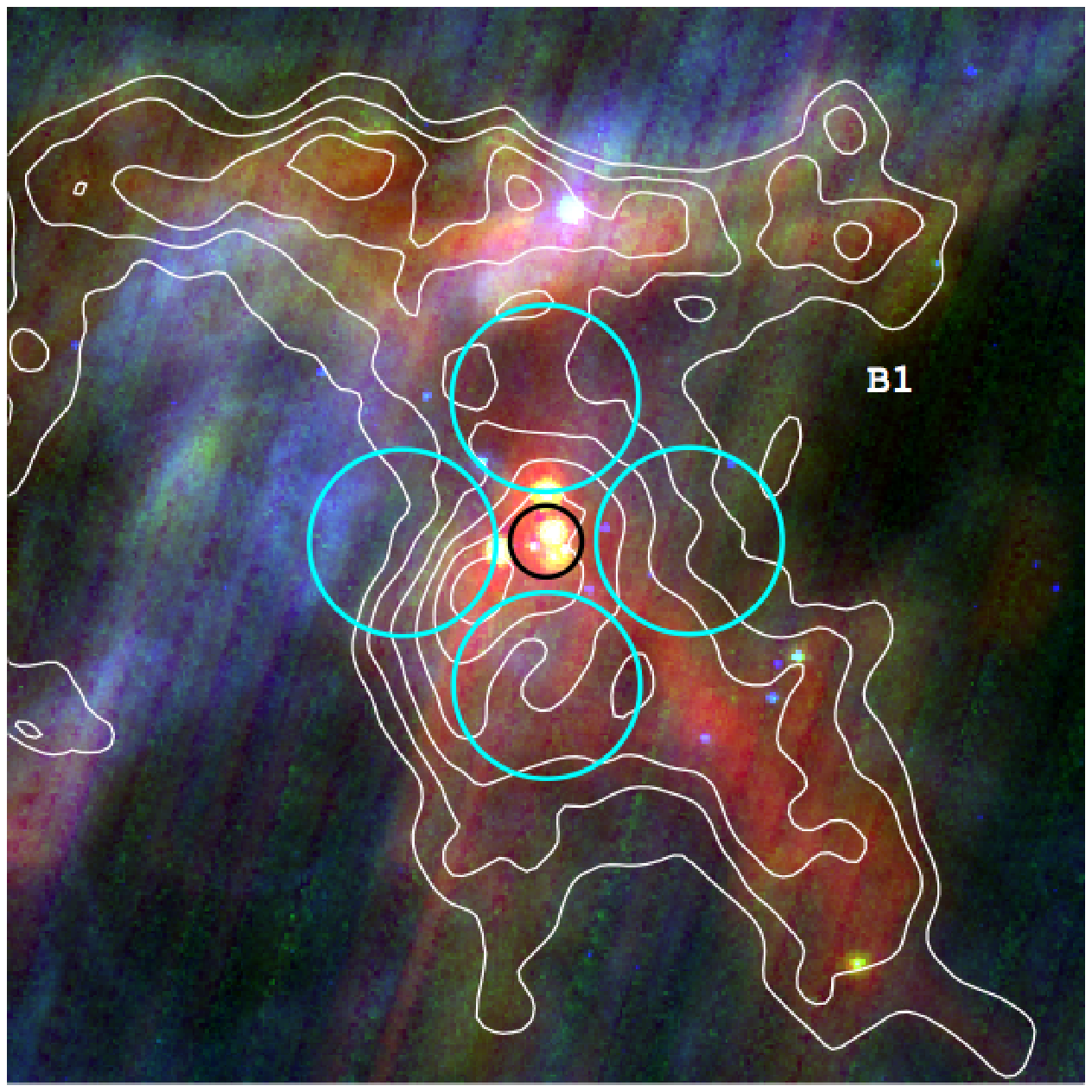}
\end{center}
\caption{Intensity maps of Spitzer continuum emission (RGB: 160/70/24 $\mu$m, respectively) and $^{13}$CO (white contours) for cores L1448CO ({\it left}) and B1 ({\it right}). The Arecibo beams and the four GBT beams that CHT09 used in each cloud are shown as black and cyan circles, respectively. Spitzer data are from the Spitzer c2d legacy program (Rebull et al.\ 2007).  $^{13}$CO data are from Ridge et al.\ (2006); contours start at 10$\sigma$ and are spaced in steps of 3$\sigma$.}
\label{perseus}
\end{figure*}

\begin{figure*}
\begin{center}
\includegraphics[width=6.5cm]{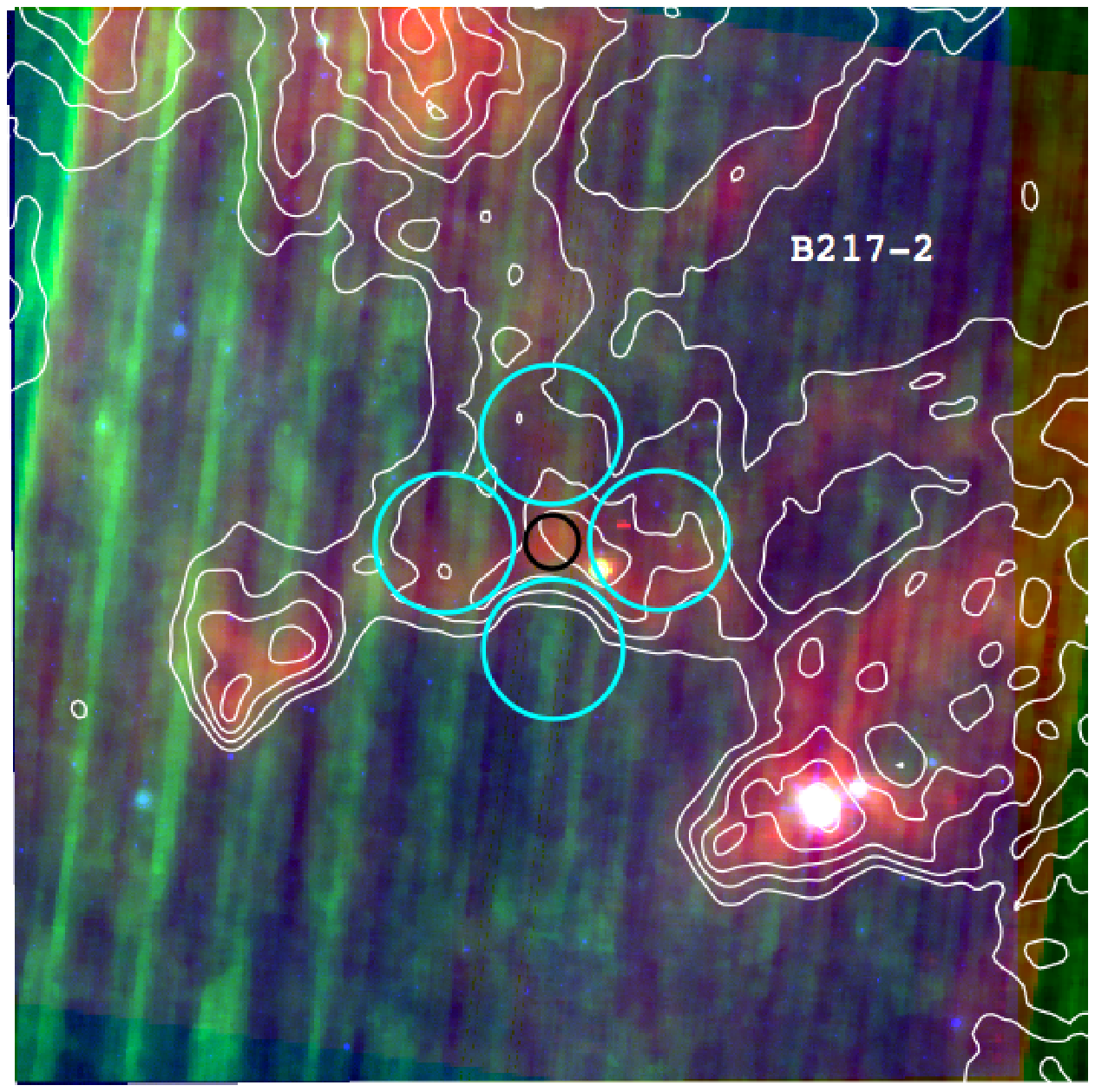}
\includegraphics[width=6.5cm]{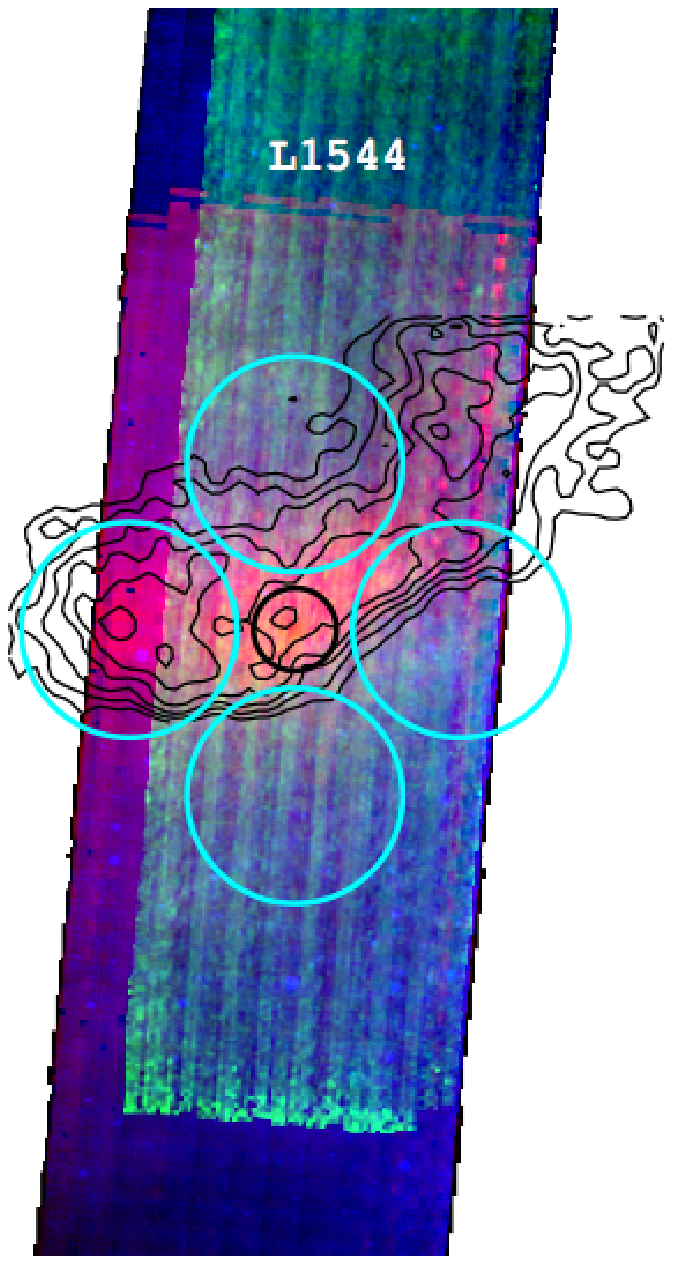}
\end{center}
\caption{As in Fig. \ref{perseus}, for cores B217-2 (left) and L1544 (right). The single detection of the envelope $B$-field in the CHT data corresponds to the GBT beam on the left (west) on the right-hand panel, where the gas density is also high. Spitzer data are from the Spitzer Taurus legacy program (B217-2) and Spitzer program \#30384 (L1544). $^{13}$CO data for B217-2 are from Goldsmith et al.\ (2008); contours start at 10$\sigma$ and are spaced in steps of 3$\sigma$.$^{13}$CO data for L1544 are from Goldsmith \& Li (2005); contours start at 3$\sigma$ and are spaced in steps of 1$\sigma$.}
\label{taurus}
\end{figure*}

Our main objection to the CHT09 analysis lies in the fact that they impose the overly restrictive {\em assumption} of zero spread in the envelope $B$-field values in each of the four clouds, without any evidence that the spread is indeed zero. (Their own data show a preference for an intrinsic spread approximately equal to the scatter induced by their (large) observational uncertainties -- see below.) It is therefore relevant to ask whether there is any {\em independent} observational evidence supporting or contradicting the assumption of no-spread in the envelope $B$-field values. To answer this question we plot intensity maps of Spitzer continuum emission (in 160, 70, and 24 $\mu$m, RGB colors) and $^{13}$CO emission (white contours), tracing the column density in the cores and their environments, in Fig. \ref{perseus} for L1448CO and B1, and in Fig. \ref{taurus} for B217-2 and L1544, respectively (see figure captions for $^{13}$CO and Spitzer data references). On each map we overplot the Arecibo beam (black circle) used for the measurements of the core magnetic field strengths, and the four GBT beams (cyan circles) used by CHT09 for the envelope magnetic-field measurements.

It is clear from these maps that the four GBT beams probe regions in the clouds' envelopes with very diverse morphologies and densities. The {\em a priori} expectation for the magnetic field then is to exhibit a similar diversity in morphology and magnitude. In fact, one of the CHT authors (Crutcher 2010) has recently argued that the magnetic field ($B$) scales with density ($\rho$) as $B \propto \rho^{2/3}$. We disagree with the exponent of this relation, but we predicted long ago that a positive correlation between $B$ and $\rho$ (namely, $B \propto \rho^{1/2}$) should indeed exist in self-gravitating, isothermal, magnetically supported objects (Mouschovias 1976; see also Fiedler \& Mouschovias 1993, Fig. 9c, and review by Mouschovias 1996). However, if one adopts their more sensitive $B - \rho$ scaling, there is an even stronger reason to expect significant variation of the magnetic-field strength in the cloud envelopes, since there is considerable density structure there. In other words, {\em there is neither theoretical nor observational justification for the CHT assumption that the magnetic-field strength is constant everywhere in each cloud envelope}.

\section{Fifteen upper limits and one measurement for $R$}\label{nocomb}

Our disagreement with CHT on the treatment of their data is about {\em the correct way to combine the four measurements in each envelope} and produce a single upper limit for $R$ (the ratio of core and envelope mass-to-flux ratio) in each cloud. An additional, important disagreement stems from the fact that neither CHT09 nor CHT10 quotes upper limits on the quantity $R$. Upper limits are the only appropriate way to quote the information content in the CHT09 data, since in 15 out of 16 cases the mean magnetic-field strength in each cloud envelope is consistent with zero (i.e., the measurements yielded nondetections).
\footnote{ Such upper limits are nontrivial to derive from the CHT09-quoted information because, even if the errors are Gaussian, the 3$\sigma$ upper limit is {\em not equal} to 3 times the error when the mean is not zero .} 

\begin{figure*}
\begin{center}
\includegraphics[width=6cm]{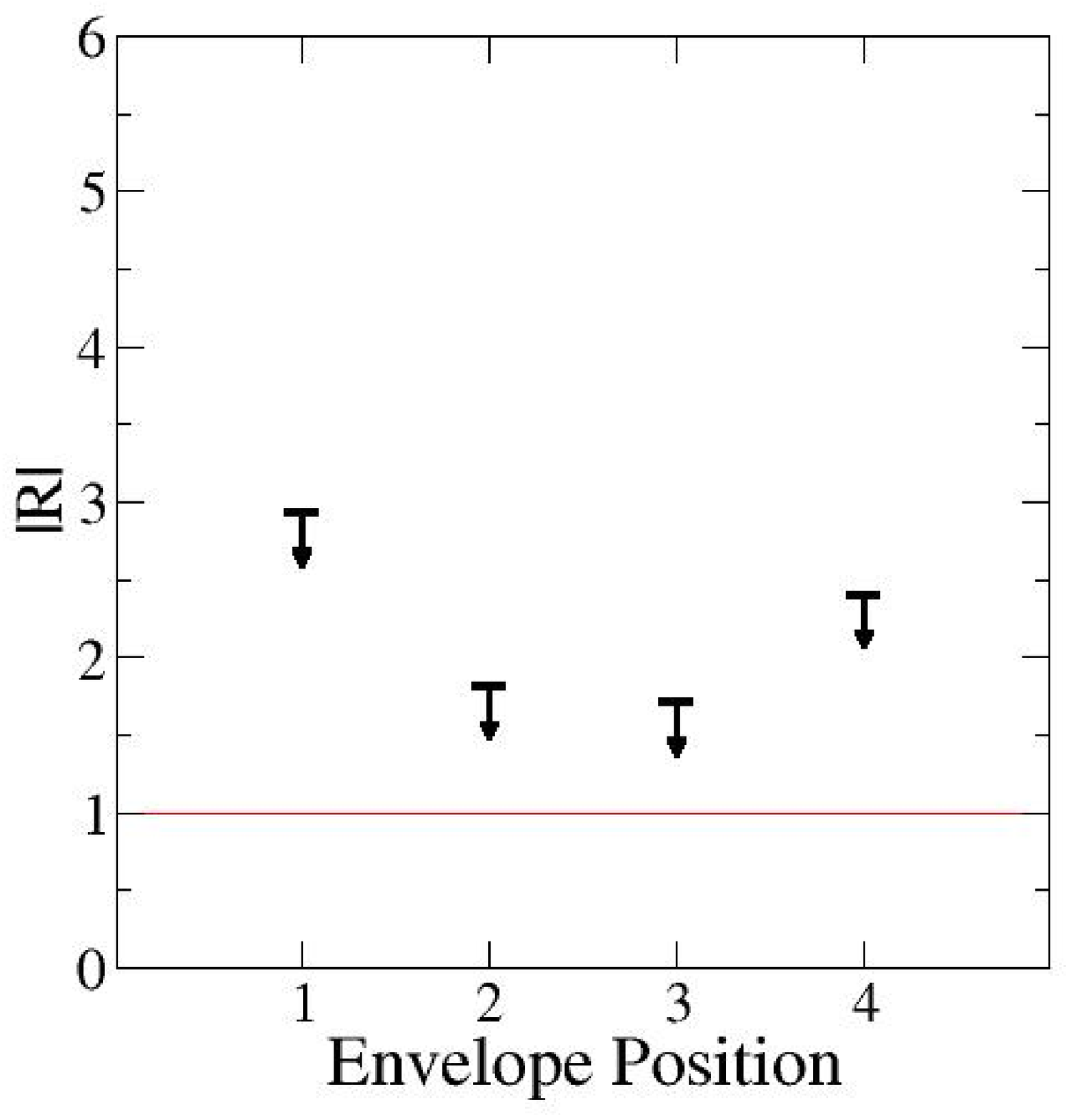}
\includegraphics[width=6cm]{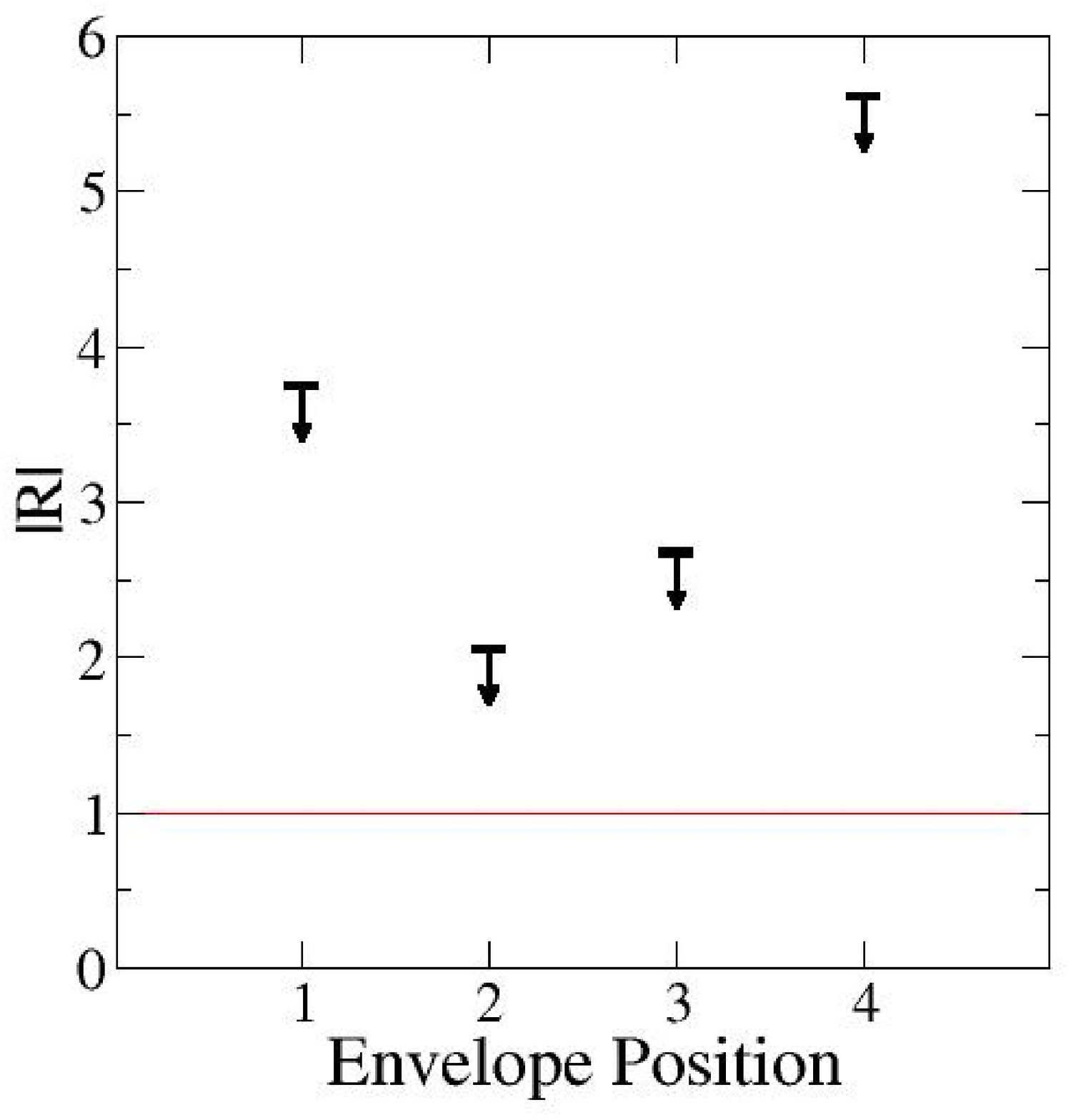}\\
\includegraphics[width=6cm]{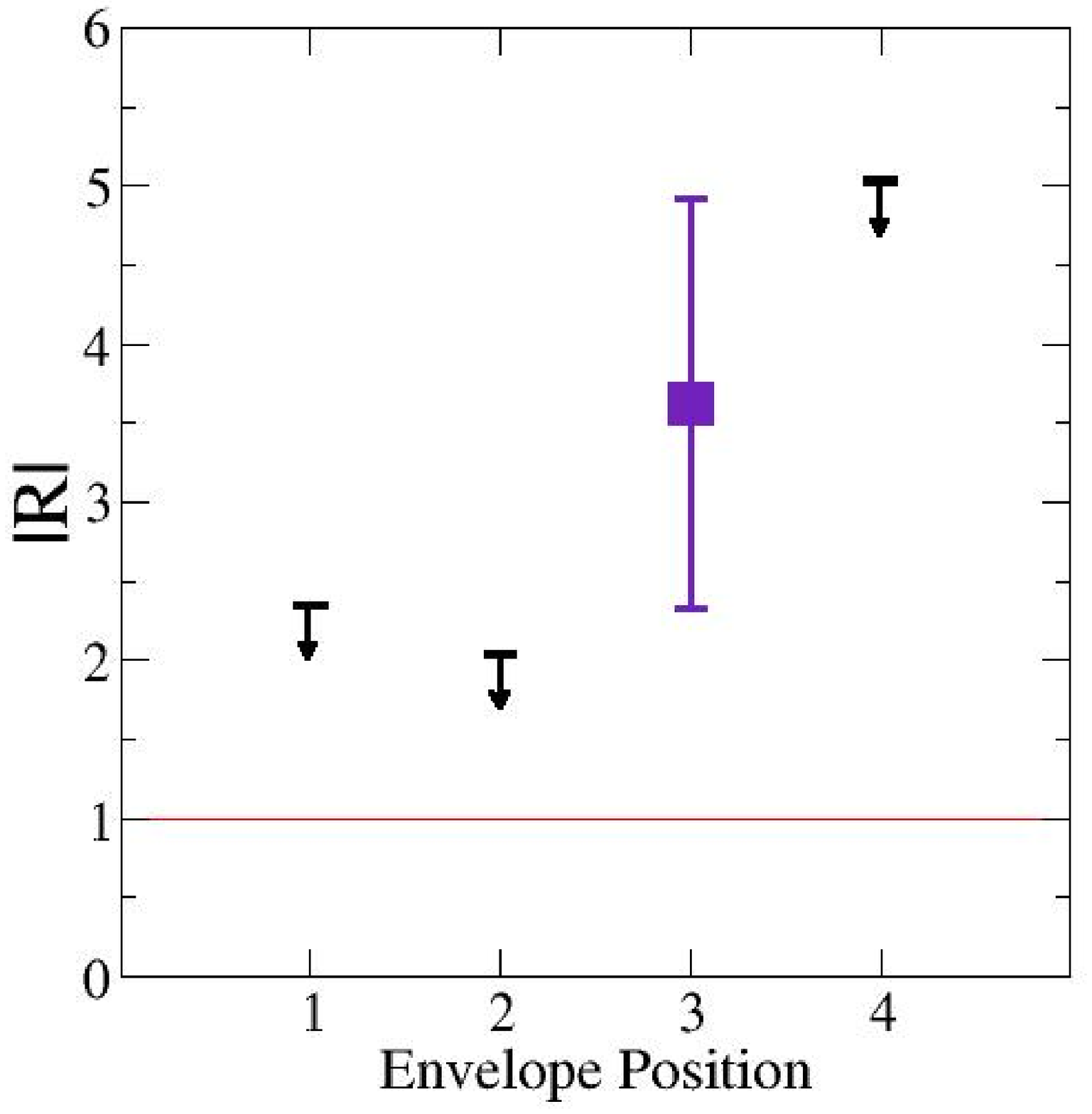}
\includegraphics[width=6cm]{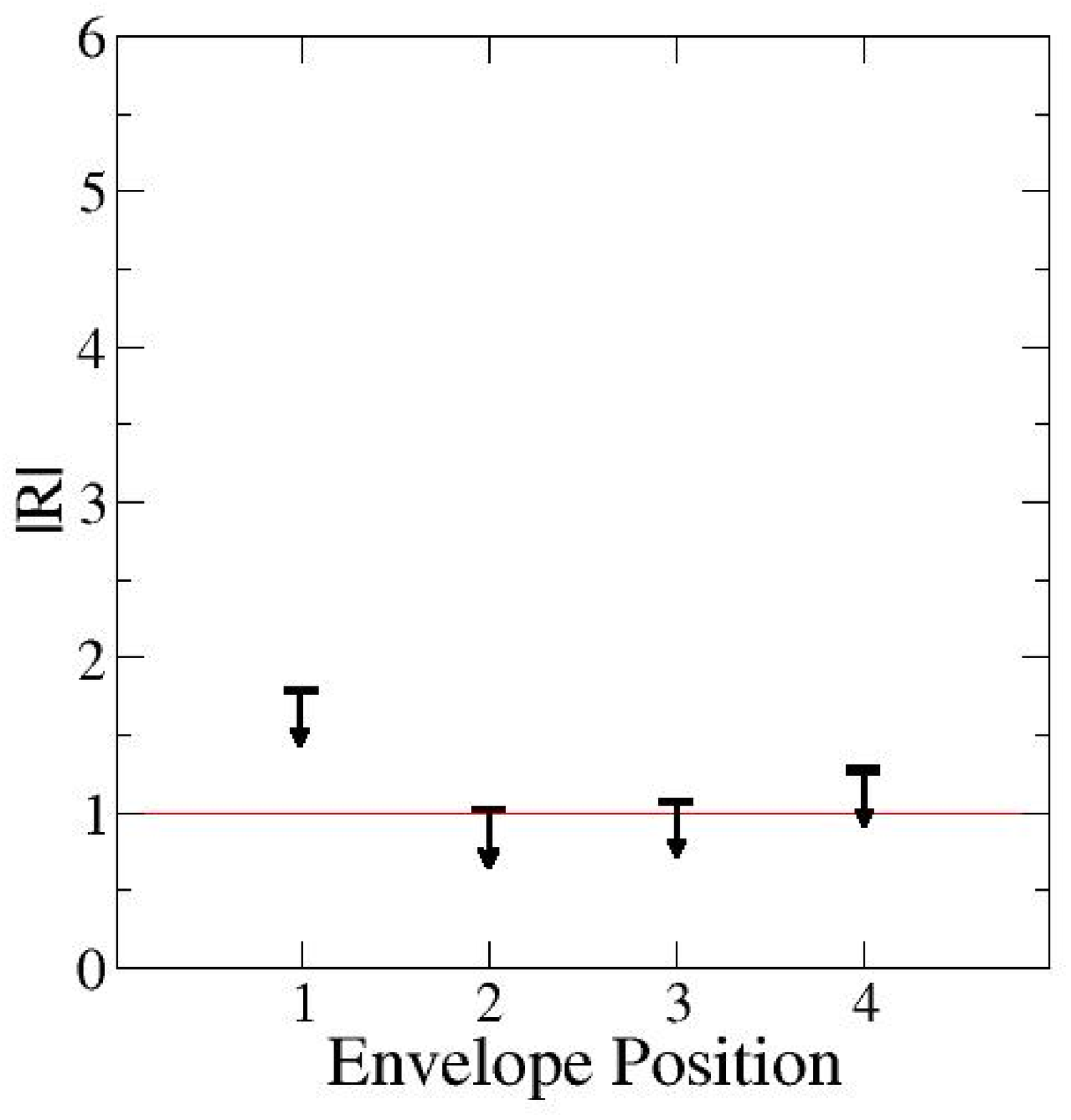}
\end{center}
\caption{3$\sigma$ upper limits (and one detection) of $|R|$ in each of the four envelope positions in each cloud; {\it upper left}: L1448CO; {\it upper right}: B217-2; {\it lower left}: L1544; {\it lower right}: B1. The square (purple) data point with errorbar is the only $3\sigma$ detection of $R$ (at the position where there is also a $3\sigma$ detection of $B$ in the envelope).  The (red) horizontal line is drawn at the value $R=1$. All 3$\sigma$ upper limits are consistent with $R>1$, and the one $3\sigma$ detection yields $R=3.5 \pm 1$. No legitimate combination of these data can yield the CHT09 value $R = 0.02 - 0.42$. Any manipulation of these data that gives an upper limit on $R < 1$ cannot but be fatally flawed.}
\label{indones}
\end{figure*}

In order to facilitate visual examination of the data {\em before} the four envelope positions are combined (which is where the CHT09/MT09 disagreement comes in play), we show in Fig. \ref{indones} the $3\sigma$ upper limit derived for $R$ {\em in each envelope position} (labeled 1, 2, 3, 4, corresponding to north, east, west, and south, respectively). For L1544west there is a $3\sigma$ detection for the envelope $B$-field, for which we can also derive a $3\sigma$ measurement of $R$. In this case, $R=3.5 \pm 1$. In all other cases, the individual $3\sigma$ upper limits are consistent with $R>1$, and in most cases they are much greater than 1. The diversity in these upper limits is another manifestation of the likely intrinsic spread in the envelope magnetic-field values (see also discussion in \S\,\ref{maps}). 

We emphasize that {\bf \it there is no disagreement between the MT09 and the CHT09 analysis in the derivation of the uncertainties shown in Fig. \ref{indones}}. It is the combination of these, {\em individually very nonconstraining measurements}, that lead to the extremely and unjustifiably strong statement in the original CHT09 work that ``The probability that all four clouds have $R > 1$ is $3 \times 10^{-7}$; our results are therefore significantly in contradiction with the hypothesis that these four cores were formed by ambipolar diffusion.'' The reason for this conclusion is the unwarranted, unjustifiable, and overly restrictive assumption that there is no variation of the magnetic-field strength within each of the four envelopes.

\section{The CHT10 claim that there is no $B$-field variation in each cloud envelope}\label{test}

CHT10 perform the following statistical test: for each cloud, they calculate all possible independent differences between envelope $B$-field measurements in different positions, $B_i-B_j$, and they normalize the result by the error on the difference, $\sigma_{ij} = \sqrt{\sigma_i^2+\sigma_j^2}$. Then, they combine the (6 for each cloud, a total of 24) values of the quantity $(B_i-B_j)/\sigma_{ij}$, they plot them in a 0.5-bin histogram, and they overplot the distribution (normal with standard deviation equal to 1) that points drawn from a single $B$-field value should obey if their spread is due to observational error only (no intrinsic spread). Their conclusion is that the two appear to be in reasonable agreement with each other. If in addition they reject three of the 24 data points associated with the L1544w measurement of the envelope $B$-field (which they characterize as ``anomalous''), the agreement between histogram and Gaussian curve appears to further improve. 

This statistical test, which is the main feature of the CHT10 paper, is claimed to prove that the magnetic field measurements in the four cloud envelopes are collectively consistent with the assumption that, in each envelope, they are drawn from a single magnetic-field value. 

The necessary condition for the CHT analysis to be valid is that a spread in the $B$-field magnitude be {\em rejected}. We show here {\em using the CHT10 test itself} that, far from that being the case, a spread in the $B$-field magnitude is in fact {\em preferred}. Moreover, we caution that there are three additional, serious problems in the CHT10 test: 
(a) CHT10 use a binned histogram to compare the data with the theoretically expected distribution; however, binning is sensitive to subjective choices, so the cumulative distribution should be used instead; (b) CHT10 exclude as ``anomalous'' the {\em only detection} of the envelope $B$-field; and (c) the stacking of all four clouds in a single statistical ``basket'' to improve the number statistics is not legitimate in this context.  

\subsection{Histogram vs Cumulative Distribution}

\begin{figure}
\begin{center}
\includegraphics[width=8cm]{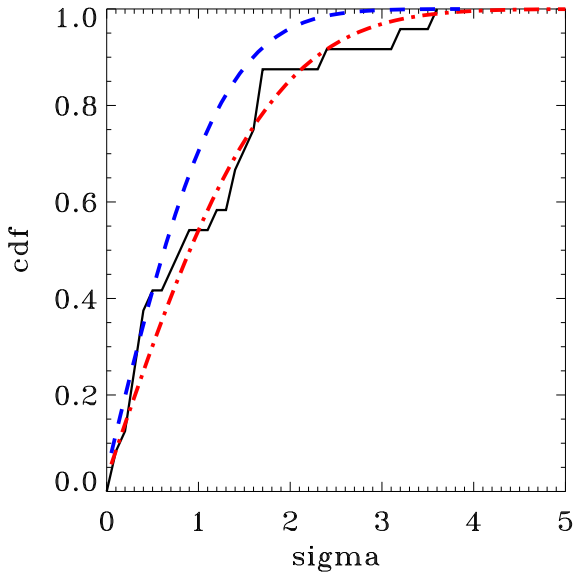}
\caption{Cumulative distribution of the quantity $(B_i-B_j)/\sigma_{ij}$. Blue (dashed line): normal distribution with standard deviation equal to 1. Red (dot-dashed) line: normal distribution with standard deviation equal to $\sqrt{2}$. The latter is a much better fit to the data, implying a spread in $B$-field values essentially equal to the observational uncertainty.}
\label{cdf}
\end{center}
\end{figure}

In the low-number--statistics limit that is applicable for the few CHT10 data points, binning choices, such as the width of the bin, can affect significantly the extent to which a distribution appears compatible with a theoretical expectation. For this reason, it is preferable to use the binning-independent {\em cumulative} distribution, which shows the fraction of points with value smaller than $x$ as a function of $x$. The cumulative distribution of the quantity $(B_i-B_j)/\sigma_{ij}$ is plotted as a solid line in Fig. \ref{cdf}. We overplot as a (blue) dashed line the cumulative distribution of a Gaussian with standard deviation equal to 1 (the distribution that, according to CHT10, corresponds to the scenario of no-spread in the envelope $B$-field). We also overplot as a (red) dot-dashed line the cumulative distribution of a Gaussian with standard deviation equal to $\sqrt{2}$ (representing essentially as much spread as observational uncertainty).

It is immediately clear from Fig. \ref{cdf} that the distribution with equal amounts of spread and observational uncertainty is a better description of the data. However, we also calculate the Kolmogorov-Smirnov test-statistic for each case, which is $t_{{\rm KS}, \sigma=1} = 0.0649$ for the $\sigma = 1$ case, and $t_{{\rm KS}, \sigma=\sqrt{2}}=0.0099$ for the $\sigma = \sqrt{2}$ case. Both distributions are allowed, because $t_{{\rm KS}}$ never exceeds the value that would reject one of the distributions even at the $80\%$ level (which, for 24 data points, is $\approx 0.2$). But the test statistic is $6.5$ times smaller for $\sigma = \sqrt{2}$ than for $\sigma = 1$, which means that the possibility of equal amounts of observational uncertainty and intrinsic spread not only is allowed, but it also is a much better fit, {\em preferred} by the data themselves.

This would be the case even if the use of the particular CHT10 test on the {\em combined} dataset including measurements for all four clouds {\it were} proper. However, we explain in  \S\,\ref{nomerge} below that such a practice is, in fact, statistically {\em improper} in this context. 

\subsection{The ``Anomalous'' Data Point is the Only CHT09 3$\sigma$ Detection!}

As mentioned is \S\,\ref{test}, first paragraph, in order to make their binned histogram look more compatible with the expected distribution, CHT10 rejected their only 3$\sigma$ detection and dealt only with the upper limits on $B$. At the expense of stating the obvious, this is an invalid procedure by any scientific measure.

\subsection{Inappropriateness of Combining the Sixteen Data Points}\label{nomerge}

The statistical analysis combining the four $B$-field measurements and producing an $R-$value (or an upper limit on $R$) will be applied to each core individually. We have already discussed that the burden of proof is in showing that {\em spread can be rejected}. Let us consider an imaginary scenario where three cores really reject spread in the $B$-field values and one does not. In this scenario, the correct analysis for the core {\em with} spread is the likelihood analysis, while in the other three cores either the likelihood analysis or the CHT analysis may be used.

If one combines all four measurements, the ``signal'' from the single core with spread will be diluted and may no longer be detectable\footnote{Think of a fit to a straight line, with a single point deviating by many sigma; if we add more points consistent with the line, the statistical significance of that single point will die away as the value of the chi-square stays the same while the number of degrees of freedom increases.}. Thus, one might incorrectly conclude that the CHT10 analysis may be applied to all clouds. This of course is not true -- each cloud is analyzed separately (even in the original CHT09 paper), it is a different physical system with possibly different properties, and the analysis appropriate {\em for each individual cloud} has to be used. 

\section {Logical Fallacies in the CHT10 Arguments}\label{fallacies}

We first state each fallacy in the CHT10 arguments and we then explain why it is such.

\subsection{If the $B$-field observations are {\em consistent} with a single value everywhere in the envelope, the CHT09 analysis is correct.}\label{burdenofproof}

Commenting on their statistical test to assess the spread in their envelope $B$-field data, CHT10 state: ``If MT are correct, these data should show a scatter significantly greater than that imposed by the measurement uncertainties.'' 

What is implied here is that, in order for the MT09 analysis to be correct, the CHT09 data should be {\em inconsistent} with a single value (i.e., the possibility of a uniform envelope $B$-field should be rejected at some statistical significance level). This statement incorrectly places the burden of proof on the general method instead of the specific one. 

{\it The method used by MT09 (likelihood analysis for calculation of errors) is general and the result obtained is correct regardless of whether there is a significant spread in the $B$-field values or not}. The method automatically suppresses any outcome that is inconsistent with the data, and places increased weight in more probable outcomes. If the data reject spread in the envelope $B$-field values, the likelihood analysis will reject that outcome as well, and will only allow the no-spread outcome. 

By contrast, the method CHT09 used automatically {\em discards} any and every possibility that the data arise from a cloud with finite spatial variations in the envelope $B$-field, independently of what the data themselves show or suggest. As a result, {\it the CHT09 method would be correct only if there were absolutely no possibility of a spatial variation of the envelope magnetic-field strength}.

The burden of proof is then on the CHT09 method: in order to prove the applicability of the method, they have to show that their data {\em reject} scatter of $B$-field values to whatever confidence level they choose, {\em not} that their data are consistent with no scatter while allowing other possibilities as well. In the latter case, one simply cannot know whether scatter exists or not, and this uncertainty is also propagated in one's ability to measure the mean magnetic field of the envelope.  After all, in all but one of the sixteen measurements of envelope magnetic fields, the observations return nondetections, consistent with a single value (zero); if {\em consistency} with a single value were the only requirement, then nondetection of a quantity would  automatically allow one to treat it as if there were no intrinsic spread, when in fact there is no information on which to base such a decision!

\subsection{The CHT09 measurement errors of the envelope $B$-field are large, so it is not necessary to account for any intrinsic spread}

CHT10 comment on a simple example given in MT09, to demonstrate that ignoring intrinsic spread can lead to an underestimate in the calculated uncertainty of the mean field, as follows: ``MT give an example of possible measurements of 10 $\mu$G and 14 $\mu$G, each with uncertainty 0.1 $\mu$G, and note that the mean differs from each value by 2 $\mu$G, not the 0.07 $\mu$G given by propagation of errors. However, these 100  and 140 $\sigma$ examples are not germane to the CHT case of roughly 1 - 2$\sigma$ measurements.'' 

In other words, what CHT10 claim here is that, since the CHT09 uncertainties are only between a factor of 1 and 2 smaller than the mean, the uncertainty on the mean is not affected by intrinsic spread. This is  incorrect, as we demonstrate below. 
 
The simple example given by MT09 was extreme by design, in order to demonstrate unambiguously that, if one does not account for intrinsic spread in the data, {\em even if the observational uncertainties are tiny}, the error on the mean can be very large. In the extreme 100$\sigma$ measurement case one underestimates the error by a factor of $\sim 30$ by not accounting for the intrinsic spread. In the case of the $1-2 \sigma$ CHT09 {\em nondetections}, we find with our detailed analysis an underestimate of the error by a factor $\simeq 2$ by ignoring the intrinsic spread. 

This can also be trivially demonstrated with a simple example similar to the one used in MT09, but specifically tailored to the CHT09 values: Consider a cloud envelope in which the magnetic field has a distribution of values with mean equal to 10 $\mu G$ and spread $5 \, \mu G$. An observer makes only two measurements of the envelope field, each with uncertainty of $5 \, \mu G$. The distribution of observables in this case has a mean of 10 $\mu G$ and a spread of $7 \, \mu G$ (the spread and observational errors added in quadrature). The first measurement gives $3 \pm 5 \, \mu G$, and the second measurement gives $5 \pm 5 \, \mu G$ (both very likely, within 1$\sigma$ from the mean). Under the CHT09 assumption of zero spread, the mean and associated uncertainty are simply the average, $4 \, \mu G$, and the propagated overvational error, $\sigma_{\rm mean} = 3.5 \, \mu G$. However the true mean differs from the measured value by $10-4=6 \, \mu G$, and the observational uncertainty is underestimating the error by a factor of about 2, as we also found with our detailed likelihood analysis. {\it Reductio ad absurdum} is a perfectly legitimate way to demonstrate the incorrectness of an argument; it should never have been questioned by CHT10.

\subsection{In the limit that the magnetic field is uniform in the observed cloud envelopes, the CHT09 analysis {\em rejects} that the cores formed by ambipolar diffusion, and instead {\em prefers} that they formed by turbulent fragmentation.}

The CHT09 observations were designed as a means to discriminate between turbulent fragmentation and ambipolar-diffusion--induced fragmentation. The authors conclude: ``our results are therefore significantly in contradiction with the hypothesis that these four cores were formed by ambipolar diffusion. Highly super-Alfv´enic turbulent simulations yield a wide range of relative $M$/$\Phi$, but favor a ratio $R<1$, as we observe.'' This conclusion is incorrect. {\em There is no turbulent-fragmentation scenario consistent with a single value of the magnetic field in the observed cloud envelopes.} Turbulent fragmentation {\em requires} a spread in the magnitude of the magnetic field in each envelope. Hence, there are two possibilities: 

(a) {\em There is} spread of the $B$-field values in the cloud envelopes. In this case, the CHT09 test does not have any discriminatory power. Ambipolar-diffusion--induced fragmentation is allowed (see MT09 analysis), and so is turbulent fragmentation (see Lunttila et al. 2009). 

(b) There is {\em no spread} of $B$-field values in the cloud envelopes. In this case, the cores could not have formed through turbulent fragmentation (because turbulent fragmentation requires spread in the envelope field). Fragmentation by ambipolar diffusion is also not allowed (see CHT09 analysis). {\em A third theory} of core formation is required. 

Observartions such as the set proposed by CHT {\em can} discriminate between ambipolar-diffusion--initiated and turbulence-induced core formation. However, in order to do so, the possibility of a complex geometry has to be considered in analyzing and interpreting the data. For example, observations designed to reveal the possibly complex cloud geometry (such as density maps and polarization observations mapping the plane-of-the-sky component of the magnetic field) need also be incorporated in the analysis and interpretation of the observations.  

\vspace{-2ex}
\subsection {The uncertainties derived by CHT09 for the envelope mean magnetic field are not equivalent to those derived from error propagation by MT09, because CHT09 synthesized a toroidal telescope beam.} 

CHT10 also comment on the way they derived the uncertainty on the envelope mean $B$-field: ``Moreover, CHT did not average the four envelope results for each cloud and obtain the uncertainty by error propagation; they synthesized a toroidal beam to sample the envelopes and obtained the uncertainties directly from the single envelope $B_{\rm LOS}$ measurement for each cloud.'' 

The implication here is that the toroidal beam synthesis method adopted by CHT09 automatically returns the correct error of measurement, and that this error is different from the value obtained by error propagation. The second part of this claim can be refuted directly and quantitatively: the error quoted in CHT09 is almost exactly the same as the uncertainty calculated using error propagation under the assumption of no $B$-field variation in the envelope -- it is for this reason that CHT10 are now arguing against envelope $B$-field variations. 

The reason for which this is the case, and for which the first part of the CHT claim is incorrect, can be seen immediately as follows. The CHT09 method of toroidal beam synthesis consists of a fit of {\em a single magetic-field value} that best describes the data from all four beams. The uncertainty associated with this value is obtained from the error in the fit. However, the uncertainty in this single-value fit is {\em monotonically decreasing with the addition of more constraints}\footnote{This is equivalent to the fit of a horizontal line to an increasing number of points; the error in this fit is monotonically decreasing as the number of points increases and, naturally, it is given by the no-spread weighted mean error propagation formula.}. ``Synthesizing'' the toroidal beam in the CHT09 manner adds constraints (and decreases the ``synthesized'' error) without allowing for field variation. By contrast, in a true toroidal beam, the addition of observed area {\em with intrinsically varying magnetic field values} would induce additional noise in the observations and increase the uncertainty in the global fit {\em of the single toroidal beam dataset}. The difference in derived uncertainty if one allows or a priori rejects intrinsic spread in the envelope $B$-field is thus exactly the same in the error propagation and ``toroidal beam synthesis'' methods -- a result that is also verified by calculating the associated uncertainties in the two cases. 

\subsection{The CHT {\em assumption} of no $B$-variation in the envelope is valid because this is what ``idealized'' ambipolar-diffusion models use, and the CHT experiment is designed to test only those models.}

The problem with this argument is two-fold. First, there is a fundamental difference between the specific geometry of an observed cloud and the basic physical processes that determine the properties, including the appearance, of the cloud. The former can alter the observable quantities due to line-of-sight effects and additional noise due to intrinsic spread of the quantity being measured (and this is what Figs. 1a and 1b in MT09 were designed to demonstrate), without affecting the nature of the underlying physical processes that govern the evolution of the cloud. As a result, {\em it is only the geometry part of the input to the ``idealized'' model} that can be safely rejected using such observable quantities without accounting for the possibility of more complex geometry. The logic of the argument should be as follows: ``A model with simple geometry and physics X cannot describe my data of a physical system. Hence, if physics X is correct, the geometry of my physical system is not simple.''

This brings us back to the necessity of {\em rejecting the possibility of spread} in order for the CHT09 analysis to be permissible. In the symbolic language used above, this would be equivalent to the following permissible (but incorrect) argument: ``I have proven that the geometry of my system is simple'' (in the CHT09 case, ``I have {\em rejected} the possibility of spread''). ``A model with simple geometry and physics X cannot describe my data; hence, physics X does not describe my physical system''! As we discussed in \S\,\ref{test}, CHT09 not only do not reject spread in the envelope $B$-field (i.e. not only do they not show that the geometry of the system is simple), but in fact their own test shows that spread in the envelope $B$-values is prefered (i.e. a complex geometry is, independently, a better description of their physical systems). In \S\,\ref{maps} we also presented independent evidence (based on intensity maps tracing the column density) that the geometry of these systems, contrary to the CHT09 and CHT10 {\em assumptions}, is complex. 

The second problem with the CHT10 argument discussed in this section is that the assumption of uniform magnetic field in the envelope (in both magnitude and direction) is not consistently used by CHT09 throughout the observations, thereby enhancing the internal contradictions of their analysis. For example, this assumption was not used in designing the observations: If CHT09 really believed or expected that a single value of the magnetic field strength could characterize each cloud envelope, then there would be no need to observe four envelope positions in each cloud. One position would be enough, and at that position they could have spent at least four times as much actual Zeeman integration time
\footnote{CHT09 spent on average 9.5 hours of on-target GBT Zeeeman observations for each envelope position. For comparison, the past OH detections of the core magnetic field required 30 - 60 hours of integration time, with a much stronger line (Crutcher et al.\ 1993).} 
(in fact substantially more since the various overheads would also be reduced), increasing the likelihood of a $B$-field detection. Similarly, in L1544 in which {\em they do} obtain a detection in one of the envelope positions: if they had been consistently using the assumption of a single $B$-field value in the envelope, they could have used that detection to obtain an actual measurement of $R$ for this cloud -- which, as discussed in \S\,\ref{nocomb}, is greater than 1; it is 3.5 $\pm$ 1, as predicted by the ambipolar-diffusion theory. 

\section{Conclusion}

We have shown/explained that: 

\begin{enumerate}
\item[1.] The general (likelihood) analysis of the OH-Zeeman data by Mouschovias \& Tassis (2009) is not missing a cosine factor, the claim to the contrary by Crutcher et al. (2010) notwithstanding. 

\item[2.] Independent observational evidence suggests that the four cloud envelopes observed by Crutcher et al. cannot be characterized by a constant value of the magnetic-field strength, contrary to the Crutcher et al. assumption that it can be. 

\item[3.]  The Crutcher et al. (2010) statistical argument, presumed to show that a single value of the magnetic-field strength can characterize each observed cloud envelope, is incorrect; the same statistical argument actually shows that a spread in the field strength comparable to the measurement uncertainty is preferred by the data themselves. 

\item[4.]  There are numerous logical fallacies in the Crutcher et al. (2010) arguments (stated and explained in \S\,\ref{fallacies}). Their defense of the Crutcher et al. (2009) data analysis is even more flawed than their original paper.

\item[5.]  When all is said and done, the one 3$\sigma$ detection of the envelope magnetic field in L1544 yields a mass-to-flux ratio increasing from the envelope to the core by a factor of 3.5 $\pm$ 1, in agreement with the predictions of the ambipolar-diffusion theory of core formation. And all other measurements, which are nondetections, yield 3$\sigma$ upper limits on the variation of the mass-to-flux ratio from envelopes to cores in the range 1 - 5, which are also consistent with the ambipolar-diffusion theory. 

\end{enumerate}

\section{Acknowledgments}

We are grateful to Nicholas Chapman for providing Figures 1 and 2 and for invaluable discussions on observational issues. TM's work was supported in part by the National Science Foundation under grant NSF AST-07-09206 to the University of Illinois. Part of this work was carried out at the Jet Propulsion Laboratory, California Institute of Technology, under a contract with the National Aeronautics and Space Administration.  
\textsl{\textsl{}}


\begin{thebibliography}{99}

\bibitem[Crutcher et al.(1993)]{1993ApJ...407..175C} Crutcher, R.~M., 
Troland, T.~H., Goodman, A.~A., Heiles, C., Kazes, I., 
\& Myers, P.~C.\ 1993, \apj, 407, 175 

\bibitem[Crutcher et al.(2009)]{2009ApJ...692..844C} Crutcher, R.~M., 
Hakobian, N., \& Troland, T.~H.\ 2009, \apj, 692, 844 

\bibitem[Crutcher (2010)]{crutch2010} Crutcher, R.~M. 2010, ``From Stars to 
Galaxies'', J. C. Tan \& S. V. Loo eds, 3.
 
\bibitem[Crutcher et al.(2010)]{2010MNRAS.402L..64C} Crutcher, R.~M., 
Hakobian, N., \& Troland, T.~H.\ 2010, \mnras, 402, L64 

\bibitem[Fiedler \& Mouschovias(1993)]{1993ApJ...415..680F} Fiedler, R.~A., 
\& Mouschovias, T.~Ch.\ 1993, \apj, 415, 680 

\bibitem[Goldsmith \& Li(2005)]{goldsmith05} Goldsmith, P.~F., \& 
Li, D.\ 2005, \apj, 622, 938  

\bibitem[Goldsmith et al.(2008)]{goldsmith08}  Goldsmith, P.~F., Heyer, M., 
Narayanan, G., Snell, R., Li, D., \& Brunt, C.\ 2008, \apj, 680, 428 

\bibitem[Lunttila et al.(2009)]{2009ApJ...702L..37L} Lunttila, T., Padoan, 
P., Juvela, M., \& Nordlund, {\AA}.\ 2009, \apjl, 702, L37 

\bibitem[Mouschovias(1976)]{1976ApJ...207..141M} Mouschovias, T.~Ch.\ 1976, 
\apj, 207, 141 

\bibitem[Mouschovias(1996)]{1996samf.conf..479M} Mouschovias, T.~Ch.\ 1996, 
in Solar and Astrophysical Magnetohydrodynamic Flows, ed. K. Tsinganos 
(Dordrecht: Kluwer), 479 

\bibitem[Mouschovias \& Tassis(2009)]{2009MNRAS.400L..15M} Mouschovias, T.~Ch., 
\& Tassis, K.\ 2009, \mnras, 400, L15 

\bibitem[Ridge et al.(2006)]{ridge06} Ridge, N. A., et al. 2006, AJ, 131, 2921

\bibitem[Rebull et al.(2007)]{rebull07} Rebull, L.~M., et al. 2007, \apjs, 171, 447 
\end{thebibliography}
\end{document}